\documentclass{article}
\usepackage[T1]{fontenc} 
\usepackage[utf8]{inputenc} 
\usepackage{ismir,amsmath,cite,url}
\usepackage{graphicx}

\usepackage{color}

\usepackage[firstpage]{draftwatermark}
\definecolor{lightgray}{rgb}{0.9,0.9,0.9}
\definecolor{darkgray}{rgb}{0.4,0.4,0.4}
\SetWatermarkFontSize{12pt}
\SetWatermarkScale{1.1}
\SetWatermarkAngle{90}
\SetWatermarkHorCenter{202mm}
\SetWatermarkVerCenter{170mm}
\SetWatermarkColor{darkgray}
\SetWatermarkText{Late-Breaking / Demo Session Extended Abstract, ISMIR 2024 Conference}

\def\lbd{}	

\usepackage[bookmarks=false]{hyperref}

\title{Diff-MST$^\textbf{C}$ : A Mixing style Transfer Prototype for Cubase}

\multauthor
{Soumya Sai Vanka$^1$ \hspace{1.5cm} Lennart Hannink$^2$} {\bfseries{
Jean-Baptiste Rolland$^2$ \hspace{1cm} George Fazekas$^1$}\\
$^1$ Queen Mary University of London, UK\\
$^2$ Steinberg Media Technologies GmbH, Germany\\
{\tt\small s.s.vanka@qmul.ac.uk, jb.rolland@steinberg.de}}

\def\authorname{S. Vanka, L. Hannink, JB. Rolland and G. Fazekas}

\begin{document}

\maketitle
\begin{abstract}
In our demo, participants are invited to explore the \textbf{Diff-MST$^\textbf{C}$} prototype, which integrates the Diff-MST model into Steinberg's digital audio workstation (DAW), Cubase. Diff-MST, a deep learning model for mixing style transfer, forecasts mixing console parameters for tracks using a reference song. The system processes up to 20 raw tracks along with a reference song to predict mixing console parameters that can be used to create an initial mix. Users have the option to manually adjust these parameters further for greater control. In contrast to earlier deep learning systems that are limited to research ideas, Diff-MST$^C$ is a first-of-its-kind prototype integrated into a DAW. This integration facilitates mixing decisions on multitracks and lets users input context through a reference song, followed by fine-tuning of audio effects in a traditional manner.
\end{abstract}
\begin{figure}
    \centering
    \includegraphics[width =0.95\linewidth,trim={0.7cm 0.3cm 0.0cm 0.2cm}]{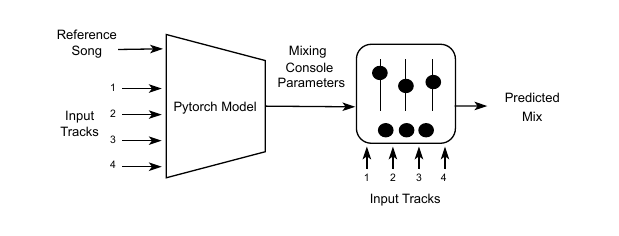}
    \caption{A High-level Overview of Diff-MST \cite{vanka2024diff}}
    \label{fig:diffmst}
\end{figure}

\section{Introduction}
\label{sec:introduction}
The democratisation of music production has introduced a spectrum of users ranging from amateurs, pro-ams, and professionals ~\cite{sai2023adoption}. Each of the user groups is skilled differently and makes use of the available technology in varied ways. Whilst amateurs want automated systems, pro-ams and professionals prefer assistive systems that are controllable and nuanced ~\cite{sai2023adoption}. The process of mixing music involves adjusting recorded music to produce an aesthetic, cohesive mix that evokes emotion. This is achieved using various audio effects like gain, pan, equalisation (EQ), compressor, reverb etc ~\cite{miller2016mixing, izhaki2017mixing}. Due to the technical and engineered nature of tools required to attain the desired artistic outcomes in the mixing process, mastering this craft requires years of training and experience.
Automatic mixing is a field of research that has explored ways for aiding, automating, and assisting in the various aspects of the mixing process ~\cite{de2019intelligent, tenyearsai}. These tools are designed to help amateurs for educational purposes and to achieve satisfactory sound quality, while also supporting professional users by streamlining technically demanding tasks and enabling faster iteration~\cite{de2019intelligent}. Beyond many classical and engineered methods explored in the past, deep learning-based approaches have shown promise in the field ~\cite{steinmetz2022automix, tenyearsai}. Deep learning-based systems for multitrack music mixing can be divided into two types: Direct Transformation (DT) systems and Parameter Estimation (PE) systems~\cite{steinmetz2022automix}. Whilst DT systems assume a black box approach offering no control, PE systems predict control parameters for the desired effect chain, allowing control and further fine-tuning. Further, the earliest systems framed multitrack mixing problem using neural networks as a supervised task~\cite{martinez2021deep, steinmetz2021automatic}, disregarding mixing fundamentally being a one-to-many problem. In real-world practice, mixing engineers receive various objects of tacit agreement like demo mixes, reference songs, and verbal descriptions for understanding the client's objective for the mix~\cite{ vanka2023role}, thus underlining the importance of context in the mixing process~\cite{lefford2021context}. Mixing engineers decide the direction for the mix based on the context that is delivered by these objects. The Diff-MST system incorporates context into the model architecture, developing on previous work ~\cite{koo2022music}.
\begin{figure*}[ht!]
    \centering
    \includegraphics[scale = 0.2]{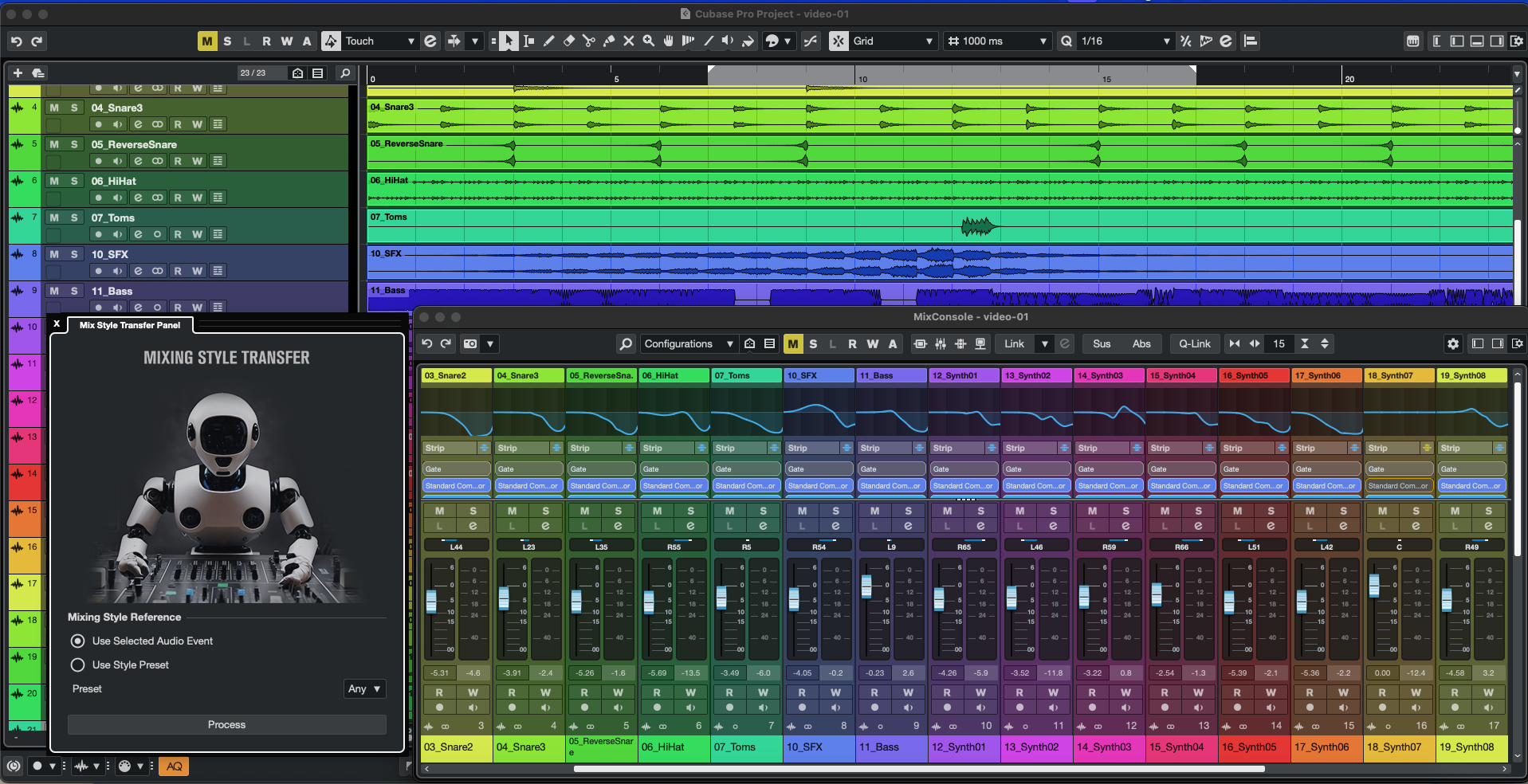}
    \caption{Diff-MSTC in Cubase}
    \label{fig:diffmstc}
\end{figure*}
\subsection{Diff-MST}
Diff-MST is a deep learning model implemented using the PE approach to conduct mixing style transfer \cite{vanka2024diff}. The system accepts multitracks and a reference song as input and provides control parameters for the mixing console along with a predicted mix styled after the reference song, as depicted in Figure \ref{fig:diffmst}. It leverages differentiable effects from the \texttt{dasp-pytorch}\footnote{\label{dasp}\url{https://github.com/csteinmetz1/dasp-pytorch/}} library, facilitating end-to-end training. The mixing console incorporates a channel strip with audio effects like gain, panorama, EQ, and compressor applied to each track. The sum of the modified outputs from all tracks is then processed through a master bus comprising an EQ, compressor, and fader. A segment of the reference song and multitrack is selected and run through the encoder. The produced embeddings are passed to a transformer encoder to obtain contextually-aware embeddings using self-attention. Finally, a controller made up of linear layers predicts the mixing console control parameters, which are fed through the mixing console with the input multitrack to produce the mix.

\section{Diff-MST$^C$}
Previous studies have demonstrated that skilled users prefer intelligent and assistive mixing systems that alleviate the technical and repetitive aspects of their work while promoting swift idea iteration \cite{sai2023adoption}. While multiple deep learning-based mixing systems have been created recently, none have been integrated into a DAW for user testing and practical application in traditional workflows. Although RoEx\footnote{\url{https://www.roexaudio.com/}} pioneered automated mixing systems, it remains web-based. Users are hesitant to send personal data to private servers, and the system's inability to run locally limits its use in studios with poor internet. Thus, there is a need for a DAW-integrated, locally run prototype. In this demonstration, we will introduce an initial prototype of a multitrack mixing system integrated into the widely-used DAW, Cubase 

\footnote{\url{https://www.steinberg.net/cubase/}}. We employed a user-centric design methodology to identify the ideal interaction and experience for the prototype. This involved an iterative process where stakeholders used brainstorming sessions to understand user needs. Ultimately, considering Cubase's existing workflows and the model's constraints, the prototype is developed to facilitate an optimal workflow. The Diff-MST model is incorporated as a Steinberg Kernel Interface (SKI) plugin within Cubase. The SKI is a private software development kit (SDK) extending virtual studio technology (VST3) capabilities. The Pytorch model is optimised for Cubase using TorchScript\footnote{\url{https://pytorch.org/docs/stable/jit.html}}. Further, we develop the user-interface for the prototype using Steinberg's VST3SDK \footnote{\url{https://github.com/steinbergmedia/vst3sdk}}.
\subsection{Workflow}
The Diff-MST$^C$ shown in Figure \ref{fig:diffmstc} includes a panel where users can either select an audio file from a muted track on their Cubase project or choose suggested songs from various genres as a reference. The user is then prompted to select a segment of the reference song. Subsequently, all tracks in the project are treated as input tracks for the model, except for those that are muted. The user must also pick a section from the project using the playback cursor to use as input for generating embeddings. The accuracy of the predicted mix largely depends on the chosen sections of the input tracks and reference song. In the final step, the user instructs the mix assistant to generate a mix. The system forecasts the control parameters for the effects. These predicted values are then applied to the effect controls of the channel strip for every active track. The user can then play back the predicted mix and make any necessary adjustments and refinements to the audio mix. The demo project can be found \href{https://youtu.be/-wXFBgWi_tw}{here}. 

\section{Conclusion}
In this demonstration, we will showcase an intelligent multitrack mixing system embedded in Cubase. This presentation supports our accepted work at ISMIR 2024 on `Diff-MST: Differentiable Mixing Style Transfer` \cite{vanka2024diff}. Participants will have the chance to try out the prototype firsthand and provide their feedback. Additionally, in the future, we'll be conducting user experience studies to scientifically evaluate the prototype and invite attendees to sign up. This work aims to bridge the gap in academic research regarding controllability of DAW-integrated intelligent mixing systems. Bridging this gap will offer valuable insights for enhancing the design of mixing systems to better cater to user needs. Moreover, this will enable us to assess the human impact of these systems and ultimately develop human-centered solutions.
\section{Acknowledgement}
This work is funded and supported by UK Research and Innovation [grant number EP/S022694/1] and Steinberg Media Technologies GmbH under the AI and Music Centre for Doctoral Training (AIM-CDT) at the Centre for Digital Music, Queen Mary University of London, London, UK.  
\bibliography{ISMIRtemplate}

%
%
%
%
%

\end{document}